\documentclass[10pt,preprint2]{aastex}
\usepackage{epsfig}

\newcommand{\msun}{\rm\,M_\odot}

\shorttitle{Nuclear Spirals as an Indicator of SMBH}
\shortauthors{Ann \& Lee}

\begin{document}

\title{NUCLEAR SPIRALS AS SIGNATURES OF SUPERMASSIVE BLACK HOLES}

\author{Hong Bae Ann \altaffilmark{1} and Hyung Mok Lee \altaffilmark{2}}
\altaffiltext{1}{Division of Science Education, Pusan National
University, Busan 609-735, Korea}

\altaffiltext{2}{Astronomy Program, School of Earth and
Environmental Sciences, Seoul National University, Seoul 151-742,
Korea, { \it e-mail}: {\tt hmlee@astro.snu.ac.kr}}

\begin{abstract}
Recent high resolution images of spiral galaxies show wide varieties of features including
nuclear spirals in the central parts. Some of the galaxies show grand-design nuclear
spirals. The morphology of
grand-design spirals can be further divided by the openness of the arms: tightly wound
ones with winding angle of around 3$\pi$ radian and open spirals with winding angle
of around $\pi$ radian. Based on hydrodynamical simulations, we have investigated the
mechanism responsible for the openness of nuclear spirals. Since the gas flow in the
nuclear region is mainly governed by the central mass concentration near the
nuclei and the sound speed of the gas, we have examined various models with
different mass concentration represented by the mass of the central black hole
and different sound speeds. We found that the tightly wound spirals
can be formed when the
mass of the black hole is large enough to remove the inner-inner
Lindblad resonances and sound
speeds lie between 15 - 20 km/sec. Thus, the presence of the tightly wound
nuclear spiral
could imply the presence of relatively massive black hole in the center.
\end{abstract}

\section{Introduction}

Supermassive black holes (SMBHs) are thought to be ubiquitous in
galactic nuclei, including the Milky Way (Genzel et al. 2003;
Gebhardt et al. 2003).
The mass of the  SMBH is found to be correlated with
the mass of the hot component of the host galaxy, with the largest mass being
a few times $10^9 \msun$ (Kormendy \& Richstone 1995; Magorrian et al. 1998).
The most
direct evidence of the presence of SMBHs at the nuclei of galaxies is the
high circular velocities or velocity dispersions of stars and gas in the
very vicinity of the
galactic centers (Marel \& Roeland 1994; Ferrarese, Ford, and Jaffe 1996).
However, most of the kinematical observations are
confined to nearby galaxies whose distances are not greater than about
10 Mpc due to the limitation of light gathering power and spatial
resolutions available for the present day ground- and space-based
telescopes.  This is the reason why the number of galaxies whose
SMBHs are identified by spectroscopic observation is quite limited.

Recent high resolution imaging surveys of galaxies show a variety of
features such as nuclear rings, nuclear bars, and nuclear spirals in the
circum-nuclear regions of normal spiral galaxies (Phillips et al 1996;
Elmegreen et al 1998; Laine et al 1999; Carollo et al 2002) as well as
active galaxies (Regan \& Mulchaey 1999; Martini \& Pogge 1999;
Laine et al 2002). Nuclear bars, sometimes called secondary bars,
have attracted special attention in recent surveys of
active galaxies (Regan \& Mulchaey 1999; Martini \& Pogge 1999) since they
are thought to provide an effective mechanism for fuelling the active galactic
nuclei (Shlosman, Begelman \& Frank  1990).
But nuclear bars are found to be somewhat rare in active galaxies
as well as in normal spiral galaxies compared to the nuclear spirals (Laine
et al. 2002; Martini et al. 2003). The high
probability ($\sim 50\%$) of the presence of nuclear spirals in active
galaxies suggests that there is
some interplay between the nuclear spirals and the SMBH
since the activities of their nuclei are believed to be powered by SMBHs.

The nuclear spirals have a variety of morphology such as
grand-design spirals, one-armed spirals, flocculent spirals, and
chaotic ones (Martini et al. $2003$). The flocculent spirals have
multiple arms which are likely to be segmented into smaller arms,
while the grand-design spirals have two symmetric arms.
%A good
%example of the flocculent nuclear spirals is that observed in NGC
%2207, which is supposed to be formed by the acoustic instability
%(Elmegreen et al. 1998; Montenegro, Yuan, \& Elmegreen 1999).
The grand-design nuclear spiral is thought to be formed by the
hydrodynamic shocks, induced by the non-axis-symmetric potential
of the large scale bar (Maciejewski et al. 2002; Ann \& Thakur
2004). Maciejewski et al. (2002) has closely examined the effects
of the sound speed on the morphology near the nuclear regions, and
concluded that the spiral density waves can penetrate deep into
the nuclear region if the sound speed of the gas is sufficiently
large (~20 km/sec). The morphology far from the IILR is nearly
independent of the sound speed. In this {\it Letter} we further
examine the effects of the central potential on the morphology of
nuclear spirals in addition to the effects of the sound speed. We
are especially interested in how the existence or absence of IILR
changes the general morphology in the nuclear region. Since bars
do not rotate fast enough to avoid IILR in galaxies with
siginificant bulge components, the IILR should be removed by
strong mass concentration such as SMBHs.

The morphology of grand-design nuclear spirals can be further
divided by the openness of the arms. A  good example of the open
nuclear spirals is NGC 5427 whose nuclear spiral winds about $\pi$
radian, while that of the tightly wound nuclear spirals is of NGC
5614 whose winding angle is about $3\pi$ radian (Martini et al.
2003). Figure 1 shows the $HST$ $H$-band images of these two
nuclear spirals. Although more than half of the active galaxies
reveal spiral patterns in the nuclear regions, the tightly wound
grand-design nuclear spirals are somewhat rare ($\lesssim 10\%$).

\begin{figure}[t]
\epsfig{figure=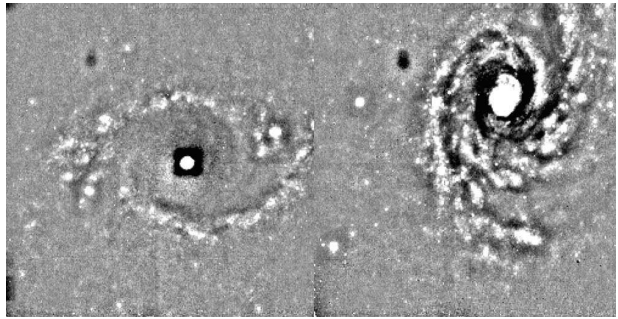, width=0.49\textwidth} \caption{Unsharp
masked images of two nuclear spirals. The left image is the
nuclear spiral of NGC 5427 which is a Seyfert II galaxy with
Hubble type of SA(s)c pec. The right image is the nuclear spiral
of NGC 5614 whose type is SA(r)ab pec. Both of the images are
taken by $HST/NICMOS$ in $H$-band. The size of each image panel is
$19.^{\prime \prime}2 \times 19.^{\prime \prime}2$. The distances
of these galaxies are 37 Mpc (NGC 5427) and 55 Mpc (NGC 5614).
These images are taken from Hubble archives.}
\end{figure}

In this letter, we investigate the physical
mechanism that is responsible for the openness of the nuclear spirals using
hydrodynamic simulations. Since the gas flow in the nuclear regions
is mainly governed by the sound speed of gas and the central mass
concentration near the galactic nuclei (Englmaier \& Shlosman 2000),
we focus on the effect of the SMBHs and the sound speeds of isothermal gas
on the morphology of grand-design nuclear spirals. We describe the models in the
next section, and provides the results of simulations in \S3. The implications
of the simulations are discussed in the final section.

\section{NUMERICAL MODELS}

We have used the smoothed particle hydrodynamic (SPH) code
incorporated with the particle mesh (PM) algorithm for the gravity
calculations in order to calculate the response of the gaseous
disk to the imposed potential. Our code is basically the same as
that of Fux (1999) but the potentials from the building blocks
(bulge, disk, halo and bar) are given by analytical functions. We
confined our simulations to two-dimensional ones to save computing
time and allow for high spatial resolutions within the disk. The
disk is assumed to be isothermal gas whose property is determined
by the sound speed. The isothermal assumption is clearly {\it ad
hoc}, but probably a plausible one since the cooling time is much
shorter than typical time scales for dynamical processes (e.g.,
Maciejewski et al. 2002). Since the gas flow depends not only on
the imposed potentials but also on the hydrodynamic properties of
the ambient gas, we vary the sound speeds of gas as well as the
mass models.
%We fixed the artificial viscosity
%coefficients as $\alpha=1.0$ and $\beta=2.5$ because the effect of
%the strength of artificial viscosity is negligibly small compared with that
%of the sound speeds.

We assumed a logarithmic potential for the dark halo, Plummer
potential for the bulge, exponential potential for the disk, and
the tri-axial potential for the bar (Long \& Murali 1992),
respectively. These potentials are easy to implement and known to
reproduce potential of the barred galaxies with suitable choices
of parameters (Lee et al. 1999; Ann \& Lee 2000; Ann 2001). We
used a point mass potential for the SMBHs with a softening radius
of 1 pc to avoid the singularity. The total mass of the galaxy
excluding the logarithmic halo is assumed to be $4\times 10^{10} ~
\rm M_\odot$. The enclosed mass of the halo within 10 kpc is
comparable to that of the visible components. In all simulations,
we assumed $R_{CR} \approx 1.2a$ where $R_{CR}$ is the corotation
radius and $a$ is semi-axis length of bar. The mass of the bar is
taken to be 20\% of the total visible mass and the axial ratio of
the bar ($a/b$) is assumed to be 3. The mass fractions and scale
lengths of the disk and bulge are selected to reproduce the
observed rotation curves of typical spiral galaxies. All the
potentials are fixed in the frame co-rotating with the bar.

\begin{figure}[t]
\epsfig{figure=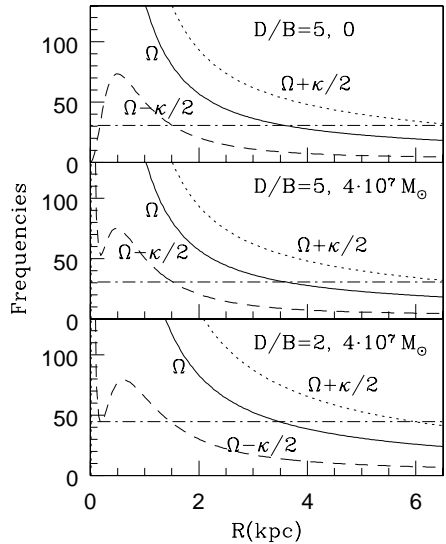, width=0.35\textwidth} \caption{Angular
frequency curves of three model galaxies. Each model consists of
three stellar components (bulge, disk and bar), dark halo and a
central black hole. The models differ in $D/B$ ratio and the mass
of central black hole, which are given in the upper right part of
each panel. The contribution of the bar component is omitted in
the plots due to it's non-axisymmetric nature. The horizontal
dot-dashed lines represent the pattern speeds of the bars which
are selected by the constraint of $R_{CR} \approx 1.2a$ where
$R_{CR}$ is the corotation radius and $a$ is the semi-major axis
of bar. }
\end{figure}

Figure 2 shows the angular frequency curves in the disk for the
models with (middle and bottom panels) and without SMBHs (top
panel). The global mass distribution for the models of top and
middle panels resembles that of late type barred spiral galaxies
whose disk-to-bulge ratio $(D/B)$ is assumed to be 5. The middle
panel shows the model with the same mass distribution as that of
the model of the top panel but with the mass of SMBH ($M_{BH}$)
being 0.1\% of the total visible mass of model galaxy, while the
bottom panel shows the model for the early type galaxies ($D/B=2$)
with the same $M_{BH}$ as that of the model of the middle panel.
As shown in Figure 2, we assumed a slowly rotating bar (pattern
speeds of 30.7 km/sec/kpc and 44.8 km/sec/kpc for early type and
late type galaxies, respectively) that allows two ILRs which are
thought to be necessary for the gas inflow to form nuclear
features such as nuclear rings and spirals in the models without
SMBH (Ann \& Lee 2000). The corotation radius lies between 3.2 and
3.7 kpc, depending on model. The presence of IILR depends on the
mass of the SMBHs as well as the global mass distribution of model
galaxies.

We have distributed  100,000 particles uniformly within a circular
disk of radius 5 kpc. Since the resolution length of SPH is mean
inter-particle distance, our simulation has initial resolution of
around 30 pc, which is small enough to study nuclear features in
detail. The resolution becomes better than the initial values in
the central parts as the density in the nuclear region increases
within a few bar rotation periods. Note that the IILR of our model
lies around 150 pc from the center.

\section{RESULTS}

We explored a wide range of parameter space for mass models and
sound speeds of gas but here we present two types of mass models
resembling early and late type barred spiral galaxies. The mass
models for early type galaxies are characterized by the
disk-to-bulge ratio of $D/B=2$, while those for the late type
galaxies are assumed to have the higher disk-to-bulge ratio (i.e.,
$D/B=5$). For each type of mass model we vary the sound speed of
gas ($c_s$) from 5 to 35 km/s and $M_{BH}$ from 0 to 1\% of the
total visible mass. We found that $c_s=15\pm5$km/s is most
plausible for the gas flow that leads to the formation of
grand-design nuclear spirals. When $M_{BH}$ is larger than $\sim
0.5\%$ of the total visible mass, the IILR is likely to disappear.

\begin{figure} [t]
\epsfig{figure=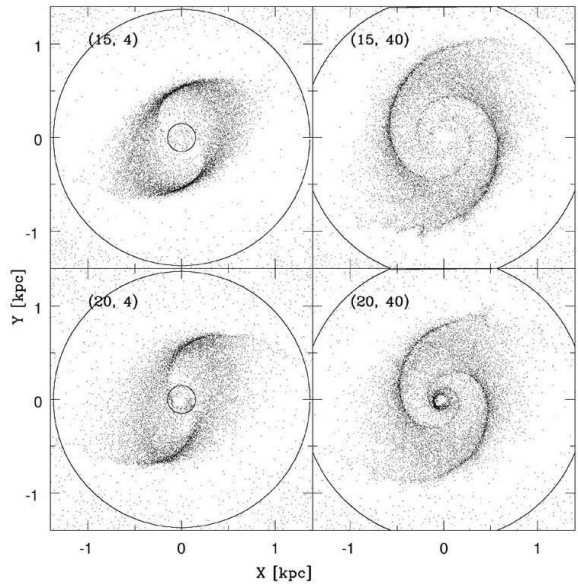, width=0.49\textwidth} \caption{Snapshots
of the nuclear region of a gaseous disk in the model galaxies that
have $D/B$=2 at the evolution time of 5 bar rotations. The models
are characterized by the sound speed of gas and the mass of SMBH
which are indicated in the upper left corner of each panel. The
units of the sound speed and black hole mass are km/s and
$10^7\msun$, respectively. The bar always lies horizontally in
these figures. The outer circles indicate the location of ILR
while the inner circles of left panels indicate the IILR. }
\end{figure}

The nuclear features of gaseous disks of the models for the early
type galaxies are shown in Figure 3.  They are the snapshots at 5
bar rotations ($\sim 0.7$ Gyr), after which the evolution becomes
very slow. The models have the same mass distributions except for
$M_{BH}$ and $c_s$. Thus, the models are characterized by the
combination of the sound speed of gas and the black hole mass,
which are indicated by the pair of numbers in the upper left
corner of each panel in units of km/s and $10^{7} \msun$,
respectively. All the models in Figure 3 clearly show spiral
features of two symmetric arms whose openness depends mainly on
the mass of SMBH. The left panels of Figure 3 resemble the open
nuclear spiral in NGC 5427, while those of the right panels show
more tightly wound spiral features. There are also some
differences in the detailed morphology of the nuclear features in
the same mass models with different sound speeds of gas. The
models with higher sound speeds of gas induce more gas inflow
close to the centers. Among the four models, the model with
$M_{BH}$ $\sim 1\%$ of the total visible mass and $c_s=15$km/s
gives the most tightly wound spiral features somewhat similar to
the nuclear spiral of NGC 5614.

\begin{figure}[t]
\epsfig{figure=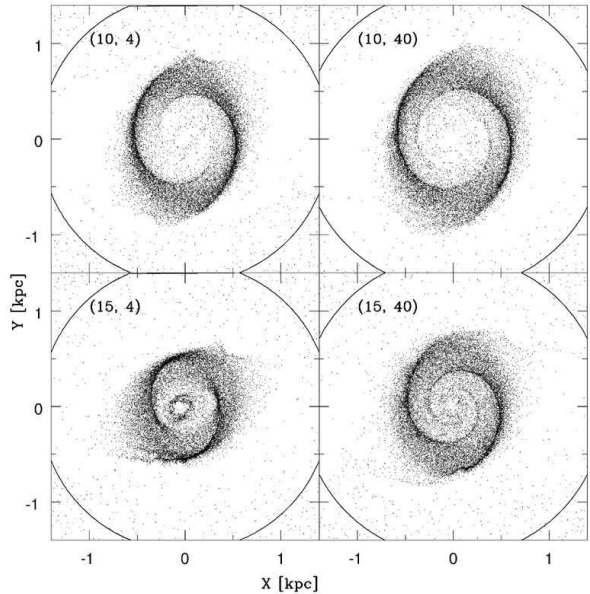, width=0.49\textwidth} \caption{Snapshots
of the nuclear regions of gaseous disks of models that represents
late type galaxies ($D/B=5$). The evolution time and meaning of
the pair of the numbers in the parenthesis are the same as those
in Fig. 3. The circles indicate the location of ILR.}
\end{figure}

The evolution of gaseous disks in the models for late type
galaxies is somewhat different from that for the early type
galaxies. Figure 4 shows the snapshots of the evolution of gaseous
disks in the models with $D/B$=5. The masses of SMBHs are the same
as those of the early type models but the sound speeds of gas are
assumed to be 10km/s and 15km/s. There seems to be no nuclear
feature similar to the open nuclear spiral of NGC 5427. Rather,
the nuclear features in the models with $c_{S}=10$km/s, shown in
the upper panels of Figure 4, show ring-like nuclear spirals
similar to the nuclear ring/spiral of NGC 4314 (Benedict et al.
1998, 2002; Ann 2001) while the model with $M_{BH}$ being 1\% of
the visible mass and $c_s$=15 km/s (lower right panel) shows a
tightly wound spiral that reaches close to the center. Similar
nuclear features are developed in the simulations that assume the
same sound speed of gas and mass model but different $M_{BH}$.
Thus, sound speed of 10km/s seems to be not high enough for the
hydrodynamic shocks to penetrate deep inside the ILRs to make
nuclear spirals there regardless of the degree of the central mass
concentration. For higher sound speed of gas ($c_{S}=15$km/s), the
models produce tightly wound spiral features when there is an SMBH
which is massive enough to remove the IILR.
%But, it is worth noting that the winding angle of the a
%nuclear spiral depends on the mass of SMBH.
%The winding angle of
%the spiral pattern in the model of $M_{BH}=4\times10^{7}\msun$
%(the lower left panel of Figure 4) is about $\pi$ radian, while
%that of the spiral pattern developed in the model with
%$M_{SMBH}=4\times10^{8}\msun$ (the lower right panel of Figure 4)
%is about $3\pi$ radian, which is very similar to that of NGC 5614.

\section{DISCUSSION}
Non-axismmetric potential by an elongated bar causes substantial
non-circular motions of stars and gas, and generates the
travelling density wave in disk galaxies. Density wave in the
gaseous disk can penetrate the ILR through the hydrodynamic shock
since the orbital speed is always much larger than the sound speed
in the disk. The loss of orbital energy by the shock causes inflow
of the gas. Our high resolution numerical simulations of gaseous
disks revealed that the morphology of the nuclear spirals depends
strongly on the presence of IILR which usually exists if there is
not much concentration of the mass in the center. The IILR can be
removed by the SMBH.

There seem to be narrow ranges for the $M_{BH}$ and the sound
speed of gas for the formation of tightly wound grand-design
nuclear spirals induced by non-axisymmetric potentials. The
primary factor that constrains the physical conditions for the
formation of tightly wound nuclear spirals is the central mass
concentration imposed by SMBH since the gas flow in the models
with relatively low $M_{BH}$ does not lead to the formation of
such  nuclear spirals. The sound speeds of gas do affect the gas
flow but the range of sound speeds plausible for the formation of
tightly wound nuclear spirals is limited by the correlation
between the gravitational potential shapes and the sound speeds in
the nuclear regions of galaxies (Englmaier \& Shlosman 2000). Our
simulations show clearly that the plausible gas sound speeds to
form the tightly wound nuclear spirals are $15\pm5$ km/s for late
type galaxies and $20\pm5$km/s for early type galaxies,
respectively.

The openness of the nuclear spirals are also known to depend on
two parameters, the central mass concentration of the host
galaxies and the sound speeds in the gas. However, as shown in
Figure 3, the openness of the nuclear spiral depends only on the
central mass concentration due to SMBH for galaxies with small
$D/B$. The gas flow in the models with large $D/B$ does not lead
to the formation of open nuclear spirals similar to that of NGC
5427. Rather, it leads to the formation of tightly wound but
ring-like nuclear spirals when the sound speed of gas is 10 km/s.
The tightly wound nuclear spirals with large winding angles are
developed in models with higher sound speed of gas ($c_s \gtrsim
15$ km/sec). Thus, the openness of the grand-design nuclear
spirals may indicate the mass of SMBH that induce the hydrodynamic
instability inside the IILR in early type galaxies, while winding
angles of the tightly wound nuclear spirals in late type galaxies
indicate the degree of the turbulent motion, represented by $c_s$
in our models, in the interstellar medium near the galactic
nuclei.

The low frequency of tightly wound grand-design nuclear spirals in
active galaxies suggests that only a small fraction of active
galaxies have SMBHs massive enough to remove the IILR. The minimum
mass of SMBH for the formation of tightly wound nuclear spirals
depends on the mass of the bulge component whose central
concentration dominates the strength of the ILRs. An SMBH having
$0.1\%$ of the total visible mass is massive enough to form
tightly wound nuclear spirals in late type galaxies, whereas the
minimum mass of SMBH to allow the formation of tightly wound
nuclear spirals in early type galaxies is about $0.5\%$ of the
total visible mass of galaxies. This does not mean that tightly
wound grand-design spirals are more frequent in late type galaxies
than early type ones because the SMBHs in late type galaxies are
likely to be smaller than those in the early type galaxies
(Magorrian et al. 1998; Ferrarese \& Merritt 2000; Gebhardt et al.
2000).

Therefore, it seems promising that the morphology of the nuclear spirals
can be used as a diagnostic for the central mass concentration as
well as the kinematic properties of the interstellar medium when it
is incorporated with the global morphology of galaxies.
More specifically, tightly wound, grand-design nuclear spirals with large
winding angle indicate the presence of an SMBH which is massive enough
to remove the IILR.

Our results are based on high resolution SPH simulations, which
clearly have some limitations. First, our simulations are purely
two-dimensional, which cannot take into account the effects of the
vertical structure of the disk. Since the thickness of the gaseous
disk is of order of a few hundreds of pc, the vertical structure
may not be important for the formation and evolution of the spiral
density waves, but this issue can only be studied in full
3-dimensional simulations. In order to keep the current
resolution, we need to have at least an order of magnitude larger
number of particles. Second, the SPH is known to have significant
numerical dissipation. This does not seem to be a serious problem
for us since we are assuming isothermal equation of state which
itself is a consequence of efficient physical dissipation. The
numerical dissipation depends on the resolution of the
simulations, but we found very little dependence of the morphology
with different number of particles.

We thank H. Kang for valuable discussion and helps in the
numerical simulations. This work is supported in part by the grant
from ARCSEC (to HBA), and in part by the KRF grant No.
2002-041-C20123 (to HML). Most of the computations are conducted
using the super computer facilities in KISTI.

\end{document}